\documentclass[superscriptaddress,aps,amsmath,amssymb,showpacs,showkeys]{revtex4-2}
\usepackage[dvips]{graphicx}
\usepackage{times}
\usepackage{braket}
\usepackage{xcolor}
\usepackage{orcidlink}
\usepackage{hyperref}
\usepackage{booktabs} 
\usepackage{subfigure}
\usepackage{siunitx}
\hypersetup{
	colorlinks=true,
	urlcolor=magenta,
	linkcolor=red,
	citecolor=blue
}
\begin{document}
	\title{Main sequence of star formation and colour bimodality considering 
		galaxy environment}
	
	\author{Pius Privatus\orcidlink{0000-0002-6981-717X}}
	\email[Email: ]{privatuspius08@gmail.com}
	\affiliation{Department of Physics, Dibrugarh University, Dibrugarh 786004, 
		Assam, India}
	\affiliation{Department of Natural Sciences, Mbeya University of Science and 
		Technology, Iyunga 53119, Mbeya, Tanzania}
	
	\author{Umananda Dev Goswami\orcidlink{0000-0003-0012-7549}}
	\email[Email: ]{umananda2@gmail.com}
	\affiliation{Department of Physics, Dibrugarh University, Dibrugarh 786004, 
		Assam, India}
	
	\begin{abstract}
		This study involves the use of friend-of-friend method on the volume 
		limited samples constructed from the Sloan Digital Sky Survey Data Release 12 
		(SDSS DR12) to classify the galaxies into isolated and non-isolated 
		environments hence to investigate the influence of the galaxy environment on 
		the main sequence of star formation, and colour bimodality. We classified 
		the galaxies into the luminous volume-limited sample with 
		$ -22.5\leq M_r \leq -20.5$ (mag), and the faint volume-limited sample with 
		$-20.5 \leq M_r \leq -18.5$ (mag). Using the WHAN diagnostic diagram we assigned 
		the  samples into star-forming, strong AGN, weak AGN, and retired galaxies
		based on their environment (isolated and non-isolated). The friend-of-friend 
		method was successful in producing consistent results regarding the stellar 
		mass-SFR and stellar mass-colour known relations. Apart from that the decrease 
		in the slope of the main sequence for star-forming galaxies by $0.04$ dex and 
		intercept by $0.39$ dex for the luminous sample was observed while the faint sample 
		a decrease of $0.08$ dex in slope and $0.74$ dex in intercept was observed 
		between isolated and non-isolated galaxies. A significant difference
		on the number of galaxies between isolated and non-isolated 
		galaxies within, above and  below the main sequence by 
		7.47\%, 28.51\%, 14.59\%
		for the luminous sample while for the
		faint sample by 16.15\%, 32.60\%, 35.23\%
		on average, respectively are observed. 
		A significant difference in the 
		number of galaxies in the blue cloud, green valley, and red sequence by 
		10.30\%, 20.61\%, 5.74\%
		for luminous sample while for faint sample by 
		28.46\%,41.36\%, 8.95\%
		on average, respectively was observed. The study 
		concludes that the galaxy environment influences the shaping and positioning 
		of galaxies along the star formation 
		main sequence and colour bimodality.
	\end{abstract}
	
	\maketitle    
	
	\section{Introduction}
	\label{secI}
	The star-forming (SF) main sequence (MS) refers to a close relationship between 
	the rates at which stars are formed and the changes in the masses of stars 
	($M_\star$s) in the majority of the galaxy population undergoing star 
	formation. The baseline of the correlation between these two parameters 
	changes over time, indicating that the intensity of star formation changes 
	significantly with time. A comprehensive understanding of how the MS evolves 
	is crucial for capturing fundamental insights into how galaxies grow. In this 
	regard, the slope of the correlation between the star formation rates (SFRs) 
	and $M_\star$ provides information on how star formation activity changes 
	across diverse stellar masses, while the dispersion of this correlation 
	unveils the degree of unpredictability in the history of gas accretion and 
	the efficiency of star formation \cite{behroozi2013average,
		primack10701924star,rodighiero2011lesser,reddy2012characteristic,
		salmon2015relation,santini2017star,speagle2014highly}. Therefore, 
	comprehending the trajectory of galaxies over time in the SFR versus 
	$M_\star$ plane serves as a potential tool for identifying the processes 
	responsible for the gradual decline in star formation activity as time 
	progresses. There is considerable uncertainty in the literature regarding the slope and 
	dispersion of this correlation.  The predominant source of uncertainty 
	arises from the biases introduced by various factors. These may include 
	criteria employed in the sample selection, evolution of gas density 
	with redshift and non-linearity of the relation, where the slope varies 
	across different mass scales.
	
	By selecting the SF galaxies from the seventh release of SDSS data \cite{abazajian2009seventh} 
	within the range of redshift $0.02 < z < 0.085$, 
	Ref.\ \cite{renzini2015objective} 
	obtained the absence of deviation from the simple power-law MS relation in 
	the local galaxies. This study revealed that the deviation originates from 
	the inconsistency in SF galaxies' selection. However, in this study, the active 
	galactic nucleus (AGN) galaxies were excluded from the analysis and only the 
	SF galaxies were discussed. Further, it is believed that AGNs might 
	be one possible cause for shifting their host galaxy's placement with respect 
	to the MS \cite{zhu2023radio}. Subsequent works by 
	Refs.~\cite{mountrichas2022comparisona,mountrichas2022comparisonb} 
	demonstrated that low-luminosity AGNs have SFRs that are below or on the MS 
	and complimented the initial finding that galaxies hosting high-luminosity 
	AGNs have enhanced SFRs compared to the MS. When studying the AGNs from MaNGA 
	survey \cite{abdurro2022seventeenth}, Ref.~\cite{riffel2023mapping} observed high-luminosity AGNs to 
	have enhanced SFRs compared to the MS.  On the other hand, 
	Ref.\ \cite{popesso2019maina} observed that using colour-colour selection 
	method of SF galaxies results in a steep MS due to the 
	exclusion of red dusty SF galaxies.
	
	The normalization of the MS evolution is thought to be due to the evolution of 
	gas density with redshift \cite{peng2010mass,speagle2014highly,
		whitaker2012star,magdis2012evolving,tomczak2016sfr,popesso2019mainb,whitaker2014constraining,
		thorne2021deep,leja2022new}. There is an observation that the MS obeys the power 
	law shape, SFR $\propto M^{\alpha}_\star$, both in the local Universe and
	distant Universe
	\cite{speagle2014highly,peng2010mass}. Other works suggest that the SFR vs M$\star$ 
	relation obeys a curvature nature towards the high mass at the low 
	\cite{popesso2019mainb} and high redshift \cite{whitaker2014constraining,popesso2019maina,leslie2020vla,thorne2021deep,leja2022new}. 
	Also, there is ongoing debate in the literature regarding the scatter around
	the relation having conflicting results \cite{leslie2015quenching,elbaz2007reversal,whitaker2012star,
		shimizu2015decreased,whitaker2014constraining,boogaard2018muse,skelton20143d,cristello2024investigating,team2020spitzer}.
	Some works, for example 
	Ref.\ \cite{elbaz2007reversal}, report a constant scatter of $0.2–0.3 $ dex 
	from low to moderate high masses. Some studies, for example, 
	\cite{boogaard2018muse}, report a decrease of scatter for the mass range  
	($10^{8}\lesssim M\star \lesssim 10^{10} /M\odot $) at different redshift.  
	Ref.\ \cite{whitaker2014constraining}, using the sample from CANDELS 
	fields \cite{skelton20143d}, within the stellar mass range 
	($8.4\lesssim M\star \lesssim 9.2 /M\odot $) at $0.5\lesssim z\lesssim 2.5$, 
	observed that the slope is dependent on stellar mass that is steeper at 
	low masses than high masses. Ref.\ \cite{cristello2024investigating}, using 
	the data from the XMM-Spitzer Extragalactic Representative Volume Survey 
	(XMM-SERVS) \cite{team2020spitzer}, observed that the less massive AGNs 
	($10^{9.5}\lesssim M\star \lesssim 10^{10.5}/M\odot $) posses 
	enhanced SFR than when compared to normal galaxies, while more massive  
	($10^{10.5}\lesssim M\star \lesssim 10^{11.5}/M\odot $) one lie on or below 
	the MS.
	
	Moreover, the galaxy colour versus stellar mass which serves as a potential 
	tool for identifying the mechanism deriving this bimodality is another very 
	an important subject to undertake 
	\cite{leslie2015quenching,schawinski2007observational}. Different 
	studies tried to include the AGNs in the colour versus stellar mass 
	diagram and obtained that these galaxies populate the red sequence and the 
	state between blue cloud and red sequence which is known as green valley. 
	For example, Ref.\ \cite{schawinski2007observational} obtained that the 
	inclusion of AGNs in these diagrams enables to obtain the role of AGNs 
	feedback in the quenching of SFR of early-type galaxies. 
	Ref.\ \cite{nandra2007aegis} obtained that the AGNs are populated on a 
	distinct region of the colour versus stellar mass diagram referred to as the 
	red sequence or top of the blue cloud, believing that a key stage in the 
	massive galaxies' evolution is the quenching of SFR at which there is the 
	migration of the blue cloud to the red sequence. On the other hand  
	Ref.\ \cite{nandra2007aegis} outlined that AGNs may be the 
	cause for the decrease of SFR as time progresses, they outlined that 
	the ongoing star 
	formation is not a necessary condition for the AGN activity since the black 
	holes' accretions are observed when the star formation has been 
	terminated.
	Ref.\ \cite{martig2009morphological} observed morphology quenching to be a 
	reason for the reduction of star formation activities in galaxies mainly as a 
	bulge formation which stabilizes the gas disc against the gravitational 
	instabilities.
	
	Different literature tried to associate the decrease of SFR with the environment, 
	for example Ref.\ \cite{belfiore2018sdss}, studying the radial profiles in 
	$H\alpha$ equivalent width and specific star formation rate (SSFR) derived 
	from spatially resolved  MaNGA  survey \cite{abdurro2022seventeenth}, 
	to gain insight into the physical
	mechanisms that suppress star formation, observed that the responsible 
	quenching mechanism appears to affect the entire galaxy. 
	Ref.\ \cite{erfanianfar2016non} obtained that morphology and environment 
	have a combined role in slowing down the star formation activities in galaxies. 
	Furthermore, they observed that a long-timescale environmental effect 
	appears at low redshift. Ref.\ \cite{lang2014bulge} suggested that the decrease 
	of SFR is mainly due to internal and linked with bulge growth. However, 
	the existence of the relation between morphology and density may provide a 
	different turn in the relation of SFR, stellar mass and the environment 
	where a particular galaxy resides. The study by 
	Ref.\ \cite{bluck2020galactic}, presenting the analysis of star formation 
	and quenching in the SDSS MaNGA survey \cite{abdurro2022seventeenth}, utilizing over 5 million 
	spaxels from $\sim 3500$ local galaxies, observed that the sudden decrease of 
	SFR affect the whole galaxy but star formation occurs in a small localized 
	scale within the galaxy. These studies aim to discern the mechanism 
	or combination of mechanisms that lead to the quenching of star formation 
	processes in galaxies, thereby influencing their evolution. 
	
	In a simplified representation, the process guiding these transformations can 
	be broadly categorized as `internal' and `external'. Rather than being driven 
	by stochastic events like massive mergers and starbursts, 
	Ref.\ \cite{speagle2014highly} demonstrated that the normalization declines 
	dramatically but steadily as a function of redshift, possibly on 
	mass-dependent time scales. Keeping in mind that more precise the 
	relation of redshift on evolution is often expressed as  
	$\propto (1 + z)^\gamma$, with $\gamma$ which range from $1.9$ to $3.7$  
	\cite{leslie2020vla,thorne2021deep,leja2022new}. Ref.\ \cite{croton2006many}, 
	using the model of galaxy formation observed the AGN feedback to be the primary 
	mechanism affecting the MS. Although this may affect the massive galaxies 
	($10^{11}$ M$\odot$), there is a lack of observed feedback effects 
	for the majority of the SF galaxies  \cite{rosario2012mean}. However, Ref.\ \cite{oemler2017star} 
	using a basic model for disk evolution, demonstrated that the evolution of galaxies away 
	from the main sequence can be attributed to the depletion of gas due to star 
	formation after a cut-off of gas inflow. This model was based on the observed 
	dependence of star formation on gas content 
	in local galaxies and assuming simple histories of cold gas inflow. 
	Ref.\ \cite{oemler2017star} further obtained that Galaxies 
	classified as MS, quiescent, or passive exhibit varying fractions 
	on mass and environment. The MS fractions decrease with increasing 
	mass and density, while the quiescent and passive fractions rise. Due to this 
	uncertainty, the ongoing debate revolves around the extent to which each of 
	these scenarios influences the shape of the relation.
	
	In this paper, our primary goal is to investigate if, the star formation MS and that of 
	colour–$M_\star$ relation depends on the 
	galaxy environment using isolated and non-isolated galaxy samples, derived 
	using modified friend-of-friend algorithms by Ref.~\cite{tempel2017merging}. 
	Our study tests if the method can produce the known facts in the literature 
	regarding the relations, and quantifying the influence of the environment on the 
	main sequence and colour bimodality.
	The width of the MS is 
	defined as $\pm\, 0.3$ dex ranging from the best-fit line of SFR–$M_\star$ 
	plane which was obtained based on the dispersion of the observed MS. 
	This approach enables us to assess the linearity of the 
	MS and explore the dispersion of the MS. Additionally, we aim to quantify 
	the evolutionary trajectory of AGN, and retired galaxies with respect 
	to the colour-stellar mass diagram, seeking insights into whether galaxy 
	environment plays a role in shaping galaxy colour bimodality. 
	
	The structure of this paper is as follows. In the next section, we explain 
	the source of data and the method of getting samples, in Section \ref{secIII} 
	we explain the methodology used in this study, Section \ref{secIV} is 
	dedicated to presenting the results. In Section \ref{secV} the results are 
	discussed. Section \ref{secVI} presents the summary and conclusion. 
	Cosmological constants are adopted from Ref.\ \cite{collaborartion2016planck}, 
	wherein the dark energy density parameter $\Omega_{\Lambda}=0.692$, Hubble 
	constant of $H_0=67.8$ km s$^{-1}$ Mpc$^{-1}$ and the matter density parameter 
	$\Omega_{m}=0.308$ are recorded.
	
	\section{Data} \label{secII}
	\subsection{The SDSS main sample}
	In this research work the catalogue data extracted from the flux-limited sample 
	of twelve releases of Sloan Digital Sky Survey (SDSS DR12) as detailed in 
	Refs.\ \cite{eisenstein2011sdss,alam2015eleventh} are used. The main 
	galaxy sample was selected from the main contiguous area of the survey-based 
	on the methods outlined in Ref.\ \cite{tempel2017merging}. 
	Galaxy data were downloaded from the SDSS Catalogue Archive Server (CAS).  
	The objects with the spectroscopic class GALAXY or QSO \cite{alam2015eleventh} was 
	selected as suggested by the SDSS team. We then filtered out galaxies with the 
	galactic-extinction-correction based on Ref.\ \cite{schlegel1998maps} where 
	Petrosian r-band magnitude fainter than $17.77$ are rejected keeping in mind 
	that the SDSS is incomplete at fainter magnitudes 
	\cite{strauss2002spectroscopic}. After correcting the redshift for the motion 
	with respect to the cosmic microwave background (CMB), using the simplified 
	formula $z_{\text{CMB}} = z_{\text{obs}} - v_p/c$, where $v_p$ is a motion 
	along the line of sight relative to the CMB, the upper distance limit at 
	$z = 0.2$ was set. The final data set contains $584449$ galaxies.
	\subsection{The volume-limited samples}
	As is already stated, the SDSS main data are flux-limited. One of 
	the disadvantage of using the flux-limited sample is that only the luminous 
	objects have a chance to be observed at large distances. Due to this problem 
	the volume-limited samples are desired and hence we constructed a volume 
	limited sample for uniformity in the sample. Due to the peculiar 
	velocities of galaxies in groups, the measured redshift (recession velocity) 
	does not give an accurate distance to a galaxy located in a group or cluster, 
	and hence to overcome this issue the apparent magnitude was transformed into 
	absolute magnitude using the relation:
	\begin{equation}
		M_{r}= m_r-25-5\log_{10} (d_L)-K,
		\label{eqR} 
	\end{equation}
	where $d_L$ is the luminosity distance, $M_{r}$ and $m_{r}$ are r-band 
	absolute and apparent magnitudes respectively, and K is the k+e-correction. 
	The k-corrections were calculated with the KCORRECT ($v4\_2$) algorithm 
	\cite{blanton2007k}. The evolution corrections were estimated using the 
	luminosity evolution model of $K_e = cz$, where $z = -1.62$ for the 
	r-filter \cite{blanton2003galaxy}. The magnitudes correspond to the 
	rest-frame (at the redshift $z = 0$) and the evolution correction was estimated 
	similarly by assuming a distance-independent luminosity function 
	\cite{tempel2012groups,tempel2014flux}.
	
	According to Ref.\  \cite{ball2006bivariatec}, the Schechter 
	function's \cite{schechter1976analytic} typical magnitude $M_r^{*}$ is around $-20.5$ mag. The 
	physical properties of galaxies have been shown to undergo an abrupt 
	transition at the typical magnitude $M_r^{*}$. The variation in the 
	clustering amplitude of galaxies with the absolute magnitude or the galaxy 
	luminosity's environmental dependency is relatively weak for galaxies fainter 
	than $M_r^{*}$, but rather strong for those brighter \cite{deng2012some}. 
	In order to better explore the galaxy properties, it is interesting to 
	compare the two types of galaxies as described above. 
	We have constructed the volume-limited samples below and above the 
	characteristic magnitude $M_r^{*}$ by calculating the effective maximum 
	distance using the relation as given by 
	\begin{equation}
		d_{\text{max}} = 10^{\left(m_{r\text{lim}} - M_{r\text{min}} + 5\right)/5} \times 10^{-6} (Mpc),
		\label{dmax}
	\end{equation}
	where $m_{r\text{lim}}=17.17$ mag, $M_{r\text{min}} =-20.5$ mag for 
	luminous and $M_{r\text{min}} =-18.5$ mag for faint samples, respectively.
	Using the $d_{\text{max}}$ 
	values and the luminosity restrictions we constructed  the luminous 
	volume-limited main galaxy sample contains $136274$ 
	galaxies with $ -22.5\leq M_r \leq -20.5$ (mag), and the faint volume-limited sample, contains 
	$26513$ galaxies with $-20.5 \leq M_r \leq -18.5$ (mag). 
	\section{Methodology}
	\label{secIII}
	\subsection{Isolated and non-isolated environment}
	For each volume-limited sample, we assigned the galaxies into isolated 
	and non-isolated sub-samples, compiled using the friend-of-friend (FoF) method 
	with a variable linking length. The essence of this approach lies in the 
	division of the sample into distinct systems through an objective and 
	automated process. This involves creating spheres with a linking length ($R$) 
	around each sample point (galaxy). To adjust the linking length based on 
	distance, the procedures outlined in  Ref.~\cite{tempel2017merging} were applied. 
	The relationship between the linking length and the redshift is represented by 
	an arctangent law as given by
	\begin{equation}
		R_{LL}(z) = R_{LL,0} \left[1 + a \arctan\left(z/z_\star\right)\right],
		\label{eq1} 
	\end{equation}
	where $R_{LL}(z)$ is the linking length used to create a sphere at a specific 
	redshift, $R_{LL,0}$ is the linking length at $z = 0$, $a$ and ${z_\star}$ are 
	free parameters. The values of $R_{LL,0}=0.34$ Mpc, $a=1.4$ and 
	${z_\star}=0.09$ are obtained by fitting equation \eqref{eq1} to the linking 
	length scaling relation. If there are other galaxies within the sphere, they 
	are considered as parts of the same system and referred to as `friends'.  
	Subsequently, additional spheres are drawn around these newly identified 
	neighbours, and the process continues with the principle that `any friend of 
	my friend is my friend'. This iterative procedure persists until no new 
	neighbours or `friends' can be added. At that point, the process concludes 
	and a system is defined. The galaxies in the system with no neighbour 
	($N\!gal = 1 $) are isolated, while the galaxies with more than one neighbour 
	($N\!gal \geq 2$) are non-isolated. Consequently, each system comprises 
	isolated galaxies or non-isolated galaxies that share at least one neighbour 
	within a distance not exceeding $R$. For the luminous volume-limited main 
	galaxy sample a total of $58038 (42.59\%)$ isolated and $78233 (57.41\%)$ 
	non-isolated galaxies were obtained, while for faint volume-limited main galaxy 
	sample a total of $10659 (40.20\%)$ isolated and $15854 (59.80\%)$ 
	non-isolated galaxies were obtained.
	\subsection{Galaxy properties}
	The stellar masses used in this study were obtained from the Max Planck 
	Institute for Astrophysics and Johns Hopkins University (MPA-JHU) team, 
	calculated from the Bayesian approach as detailed in 
	Ref.\ \cite{kauffmann2004environmental}. The Stellar mass calculation within 
	the SDSS spectroscopic fiber aperture relies on the fiber magnitudes, whereas 
	the total stellar mass is determined using the model magnitudes. 
	
	The MPA-JHU total SFR used in this study was derived from the MPA-JHU database 
	and estimated using the methods of 
	Refs.\ \cite{brinchmann2004physical,tremonti2004origin} with adjustments made 
	for the non-SF galaxies. The MPA-JHU team uses the H$\alpha$ calibration 
	\cite{kennicutt1998star} to determine the SFR for galaxies classed as SF. In 
	contrast to the approach taken by Ref.\ \cite{brinchmann2004physical}, the 
	MPA-JHU team applied aperture corrections for SFR by fitting the photometric 
	data from the outer regions of the galaxies. Specifically, for the SFR 
	computation, Ref.\ \cite{brinchmann2004physical} outlined the calculation 
	within the galaxy fiber aperture. It is important to keep in mind that the 
	region beyond the fiber SFR is estimated using the methods of 
	Ref.\ \cite{salim2015mass}. Furthermore, for the case of AGN and weak emission 
	line galaxies, SFR was determined using photometry. Keeping in mind 
	that for the case of non-SF galaxies, the ionization originates 
	from other sources, such as rejuvenation in the outer regions, post-AGB stars, 
	or ionization from AGN. As such, the SFR based on $H\alpha$ for non-SF
	galaxies need to be regarded as a maximum value \cite{sanchez2018sdss}. 
	
	\subsection{Galaxy classification}
	We classified the galaxies based on $W_{\text{H} \alpha}$ versus 
	$[\text{NII}]/\text{H}\alpha$ (WHAN) diagram \cite{cid2011comprehensive}, where 
	$W_{\text{H}\alpha}$ is the $H\alpha$ equivalent width, and $[\text{NII}]/\text{H}\alpha$ is the ratio 
	of the [NII] emission line to the $\text{H}\alpha$ line. From this diagram we 
	can have following four inequalities \cite{cid2011comprehensive}:
	\begin{align}
		\log \left( \frac{[\text{NII}]}{\text{H}\alpha} \right) < -0.4 \quad \text{and} \quad & W_{\text{H}\alpha}  > 3 \, \text{\AA},
		\label{sf}\\[8pt]
		\log \left( \frac{[\text{NII}]}{\text{H}\alpha} \right)  > -0.4 \quad \text{and} \quad & W_{\text{H}\alpha}  > 6 \, \text{\AA},
		\label{sAGN}\\[8pt]
		\log \left( \frac{[\text{NII}]}{\text{H}\alpha} \right) > -0.4 \quad \text{and} \quad & 3 \, \text{\AA} < W_{\text{H}\alpha} < 6 \, \text{\AA},
		\label{wAGN}\\[8pt]
		& W_{\text{H}\alpha} < 3 \, \text{\AA}.
		\label{RGs}
	\end{align}
	These inequalities \eqref{sf}, \eqref{sAGN}, \eqref{wAGN} and 
	\eqref{RGs} represent the pure SF galaxies, strong AGN, weak AGN and 
	retired galaxies (RGs), respectively \cite{cid2011comprehensive}. The use of 
	WHAN diagram is to increase the  accuracy of our  analysis, due to the factor that the usual traditional diagnostic 
	diagrams (BPT) of Refs.\ \cite{kewley2001theoretical,kauffmann2003host,kewley2006host, schawinski2007observational} 
	may introduce  bias  since it is well known that shock 
	ionization and AGNs could cover almost any region between the right-bottom end 
	of the loci (usually assigned to SF regions) up to the top-right 
	end of the diagram. On the other hand Low Ionization
	-Nuclear- Emission line Region (LINER)  used in BPT diagram may contain  possible  multiple 
	ionizing sources \cite{cid2011comprehensive,singh2013nature,sanchez2020spatially,sanchez2021local}.
	
\begin{figure}[h!]
		\subfigure{
			\includegraphics[width=0.42\linewidth]{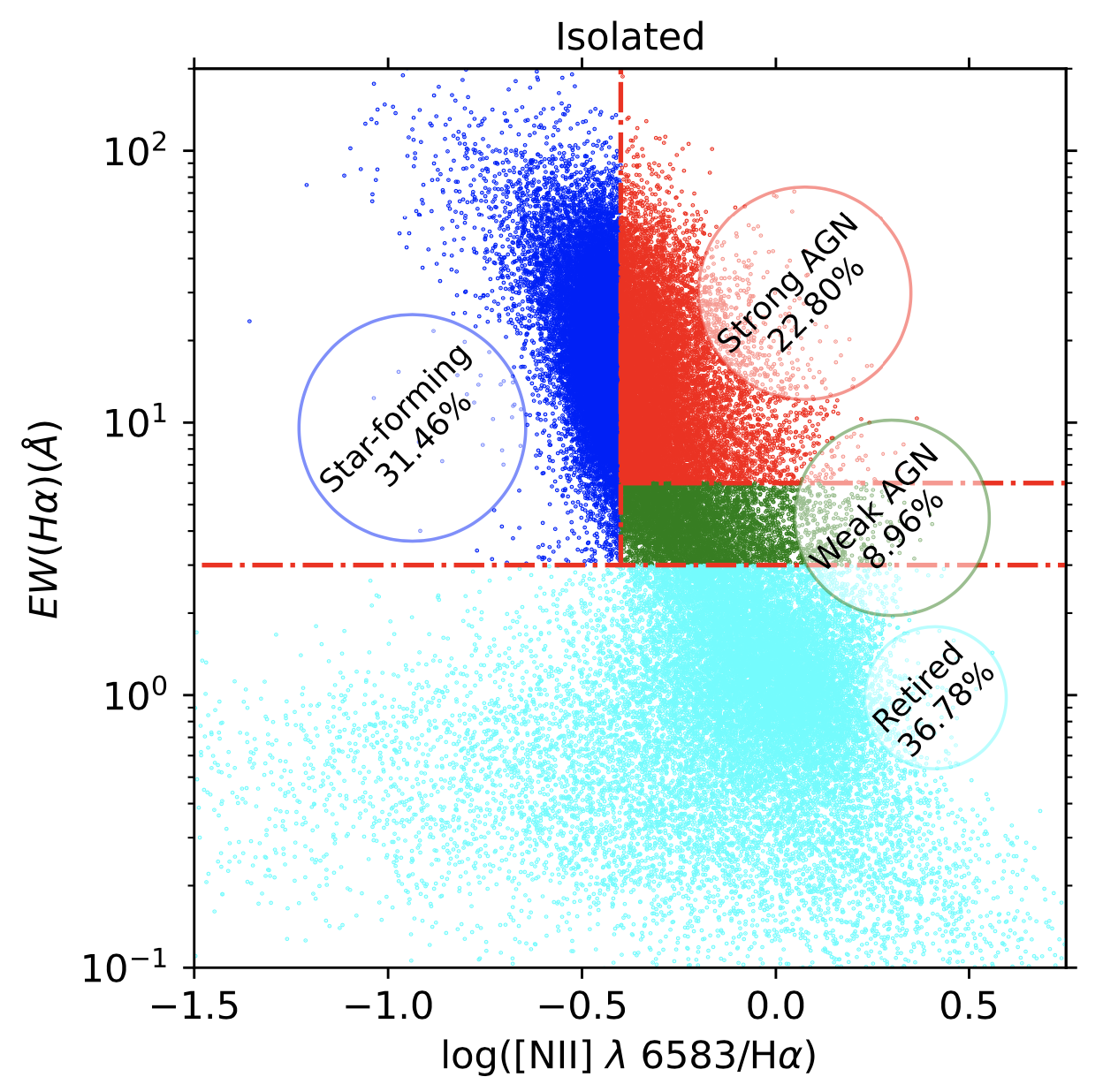}
			\label{Lisolated}
		}
		\hspace{0.3cm}
		\subfigure{
			\includegraphics[width=0.42\linewidth]{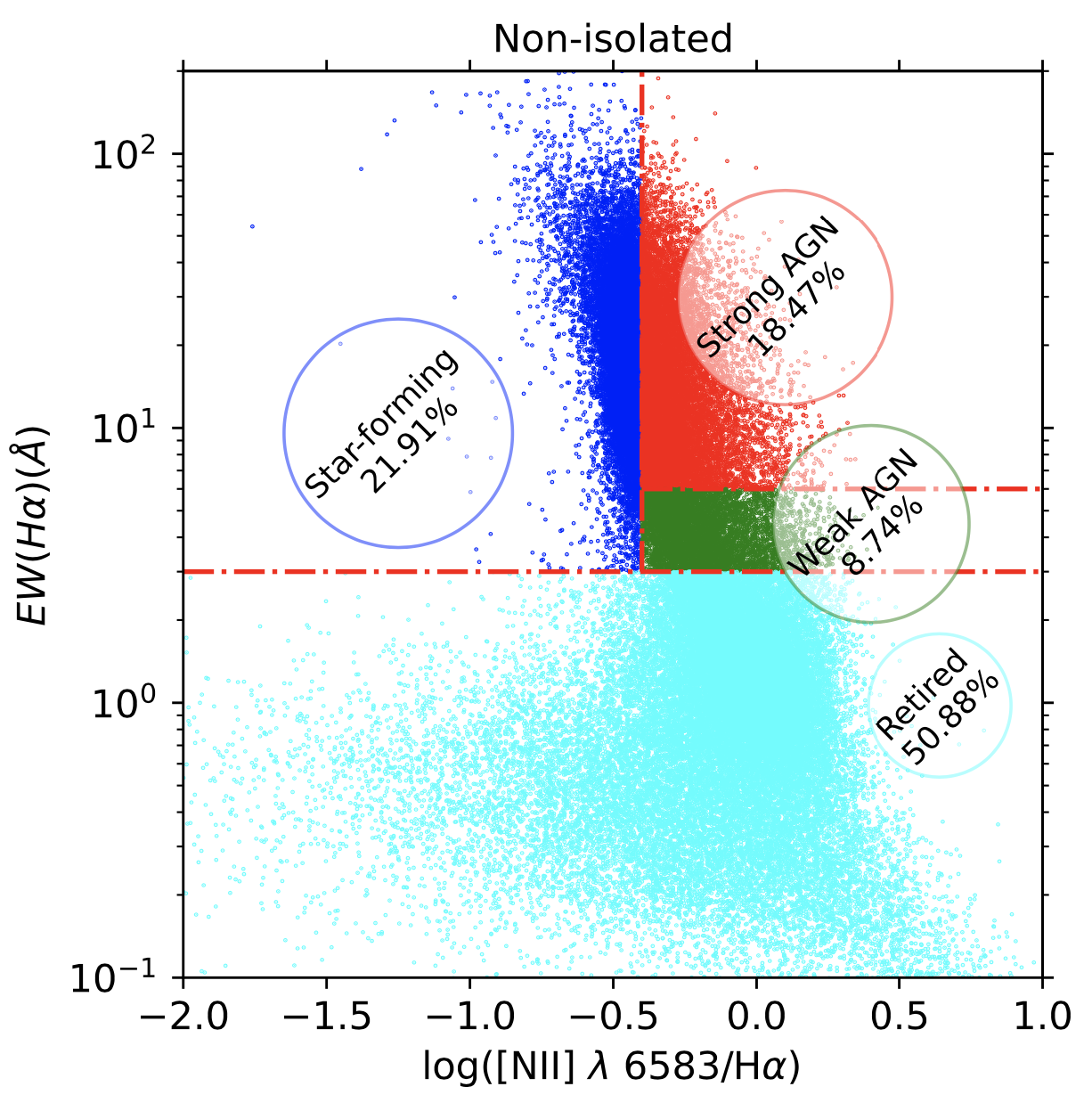}
			\label{Lnonisolated}
		}
		\subfigure{
			\includegraphics[width=0.42\linewidth]{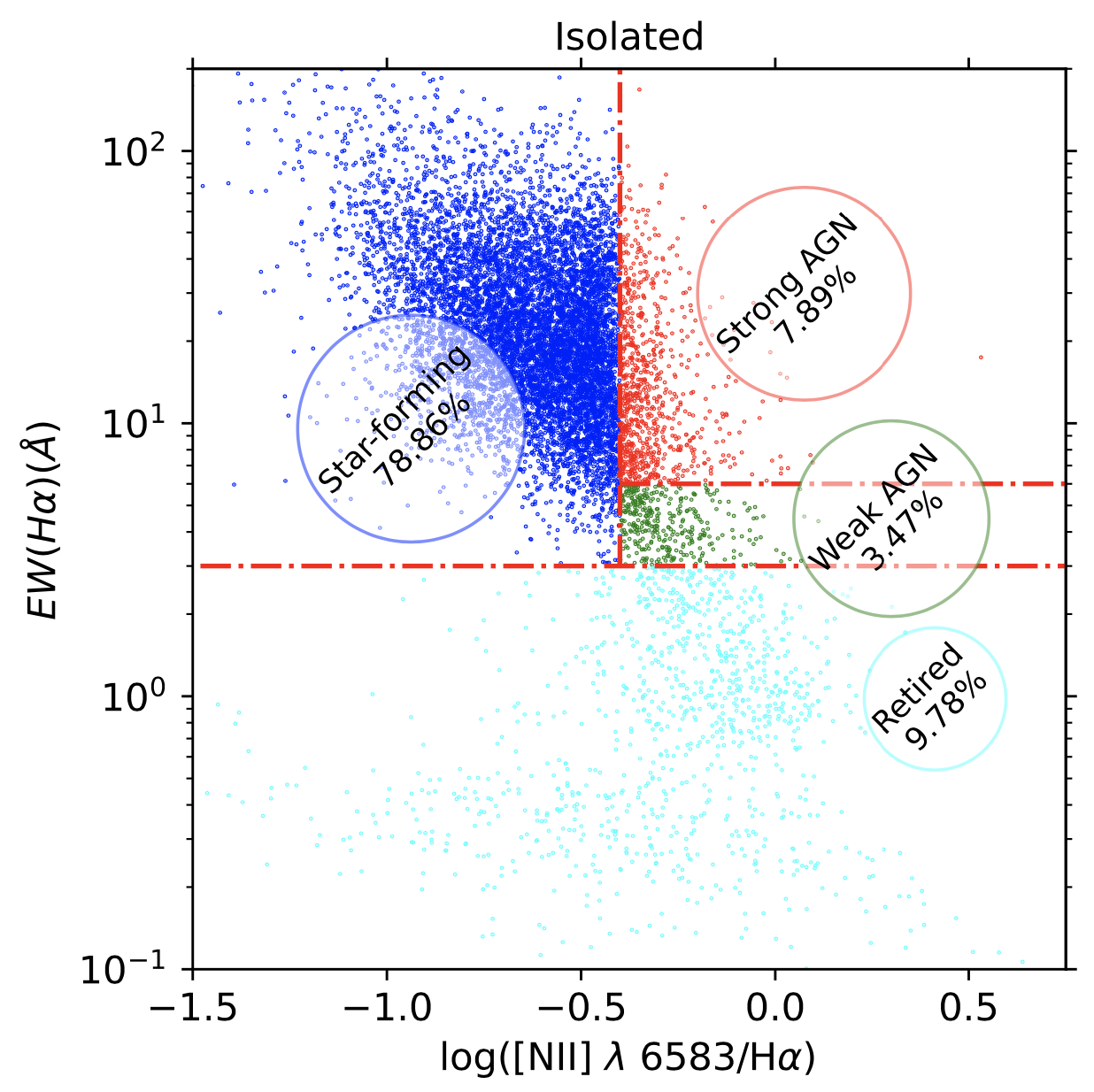}
			\label{Fisolated}
		}
		\hspace{0.3cm}
		\subfigure{
			\includegraphics[width=0.42\linewidth]{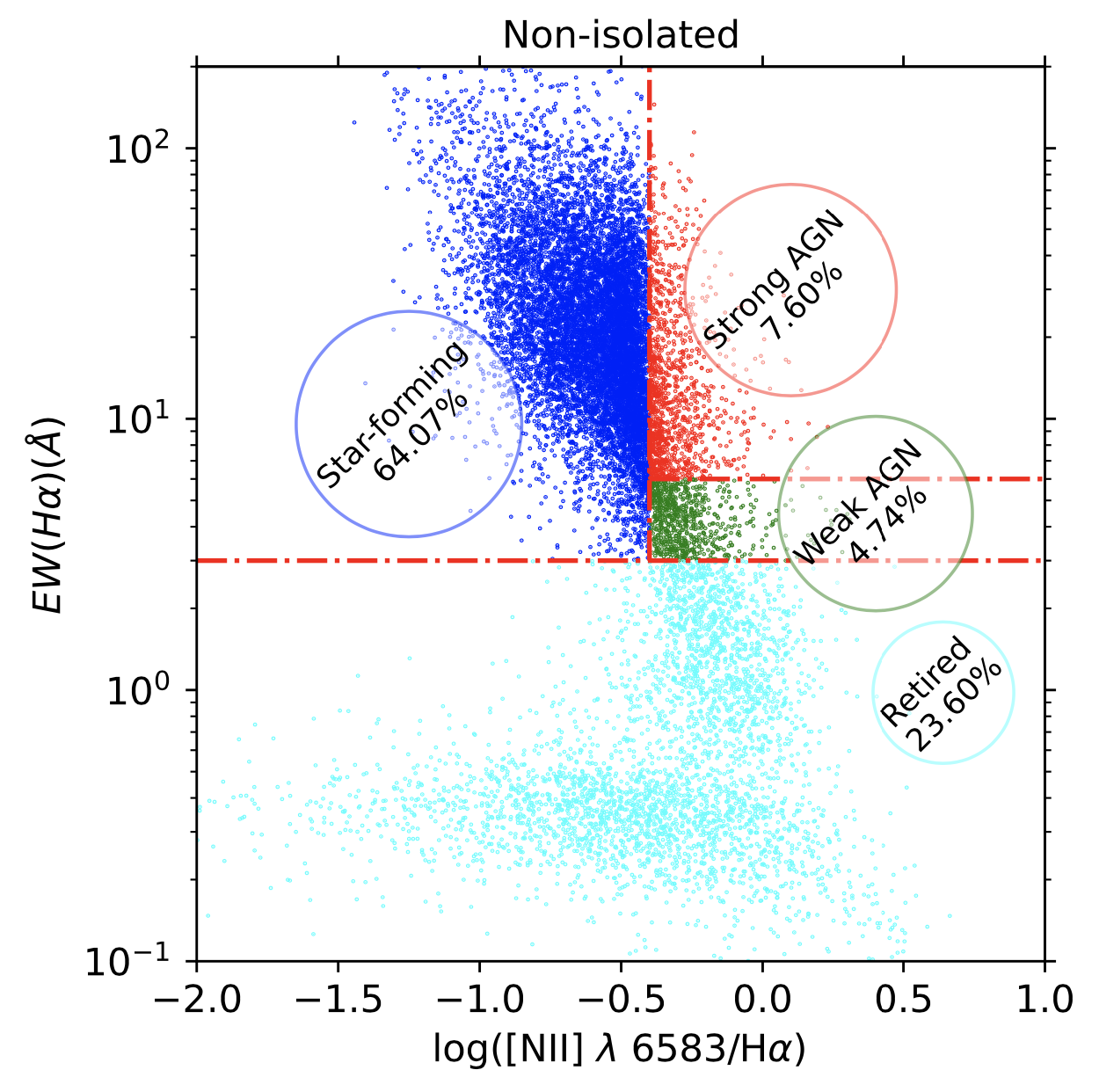}
			\label{Fnonisolated}
		}
		\vspace{-0.2cm}
		\caption{WHAN diagrams for the luminous volume limited 
			main galaxy sample (upper panels) and faint volume-limited main galaxy sample 
			(lower panels) with isolated (left panels) and non-isolated (right panels) 
			samples.}
		\label{WHAN}
	\end{figure}

	The distribution of galaxies is shown in Fig.~\ref{WHAN}, wherein the 
	SF galaxies are shown in blue colour, strong AGNs in red colour, weak AGNs in 
	green colour and RGs in cyan colour. In the luminous volume-limited sample, the following numbers are obtained, for isolated galaxies: 
	18259 ($31.46\%$) SF galaxies; 13231 ($22.80\%$) strong AGNs; 5199 ($8.96\%$) 
	weak AGNs and 21349 ($36.78\%$) RGs, while for non-isolated sample: 17145 
	($21.91\%$) SF galaxies; 14447 ($18.47\%$) strong AGNs; 6837 ($8.74\%$) weak 
	AGNs and 39807 ($50.88\%$) RGs. In the faint volume-limited sample, the 
	following numbers are obtained, for isolated galaxies: 8406 ($78.86\%$) SF 
	galaxies; 841 ($7.89\%$) strong AGNs; 370 ($3.47\%$) weak AGNs and 
	1042 ($9.78\%$) RGs, while for non-isolated sample: 10157 ($64.07\%$) SF 
	galaxies; 1205 ($7.60\%$) strong AGNs; 751 ($4.74\%$) weak AGNs and 
	3741 ($23.60\%$) RGs.
	
	\section{Results}
	\label{secIV}
	\subsection{The Main Sequence (MS)}
\begin{figure}[h!]
		\subfigure{
			\includegraphics[width=0.42\linewidth]{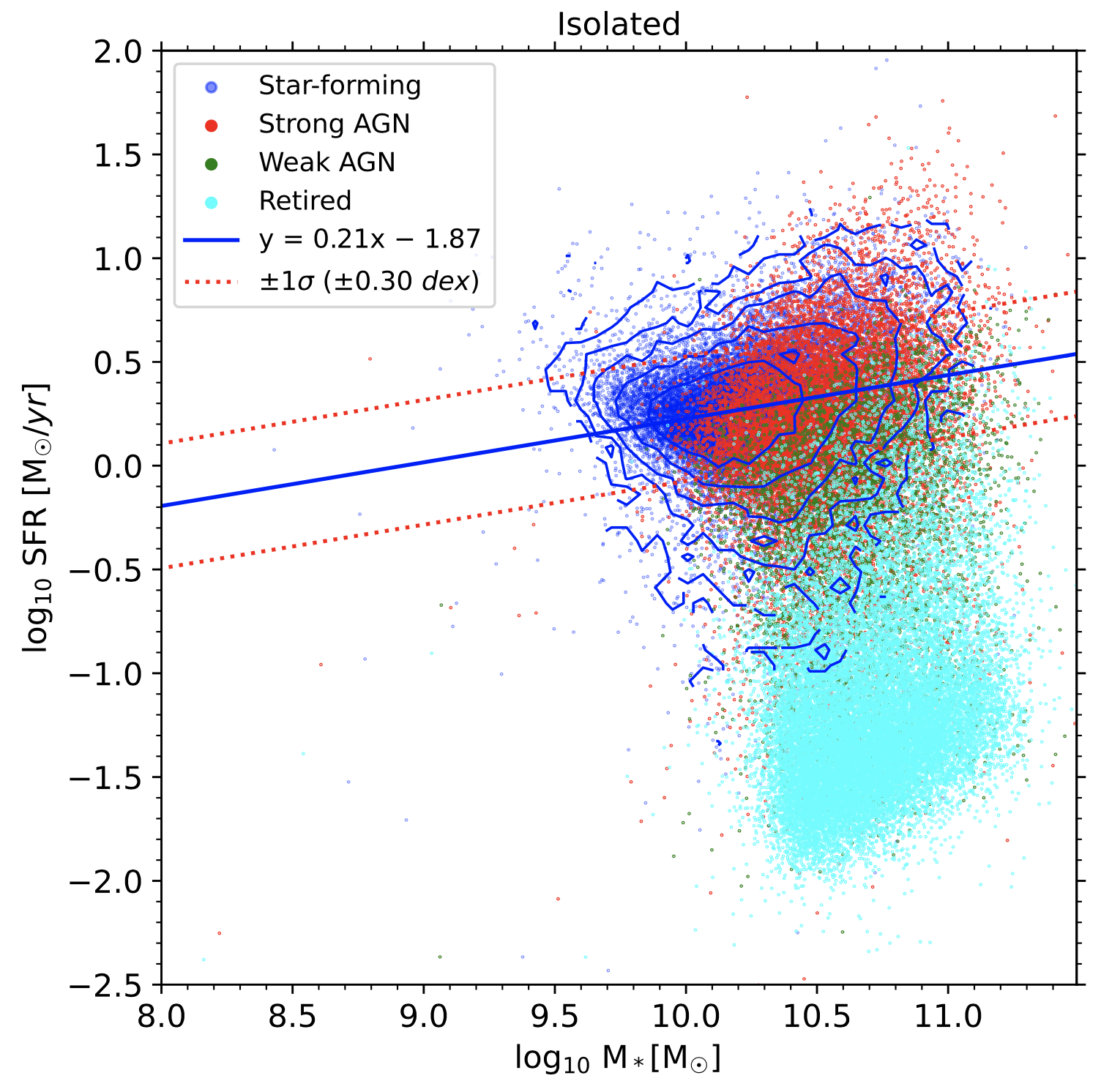}
		}
		\hspace{0.3cm}
		\subfigure{
			\includegraphics[width=0.42\linewidth]{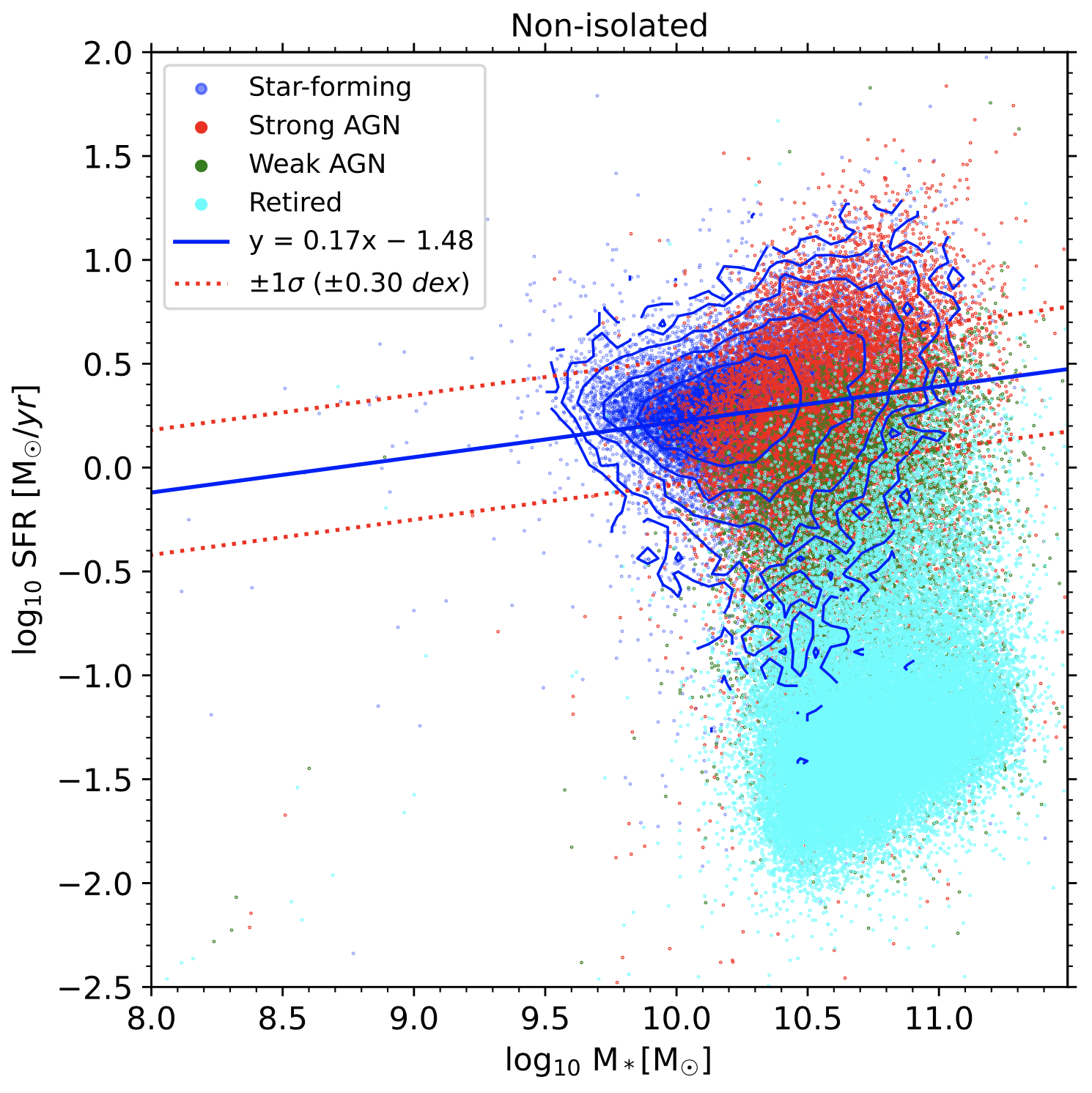}
		}
		\subfigure{
			\includegraphics[width=0.42\linewidth]{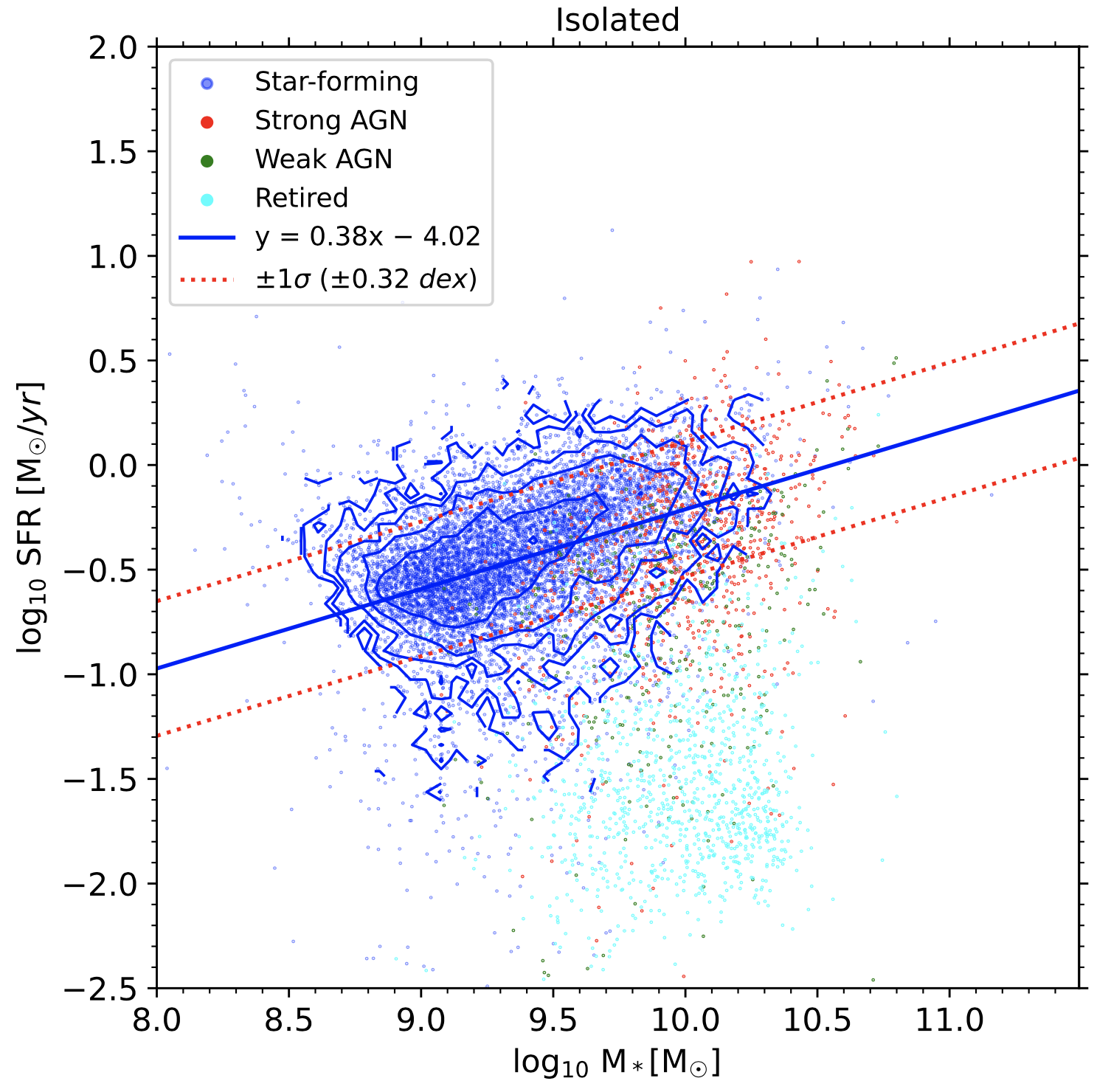}
		}
		\hspace{0.3cm}
		\subfigure{
			\includegraphics[width=0.42\linewidth]{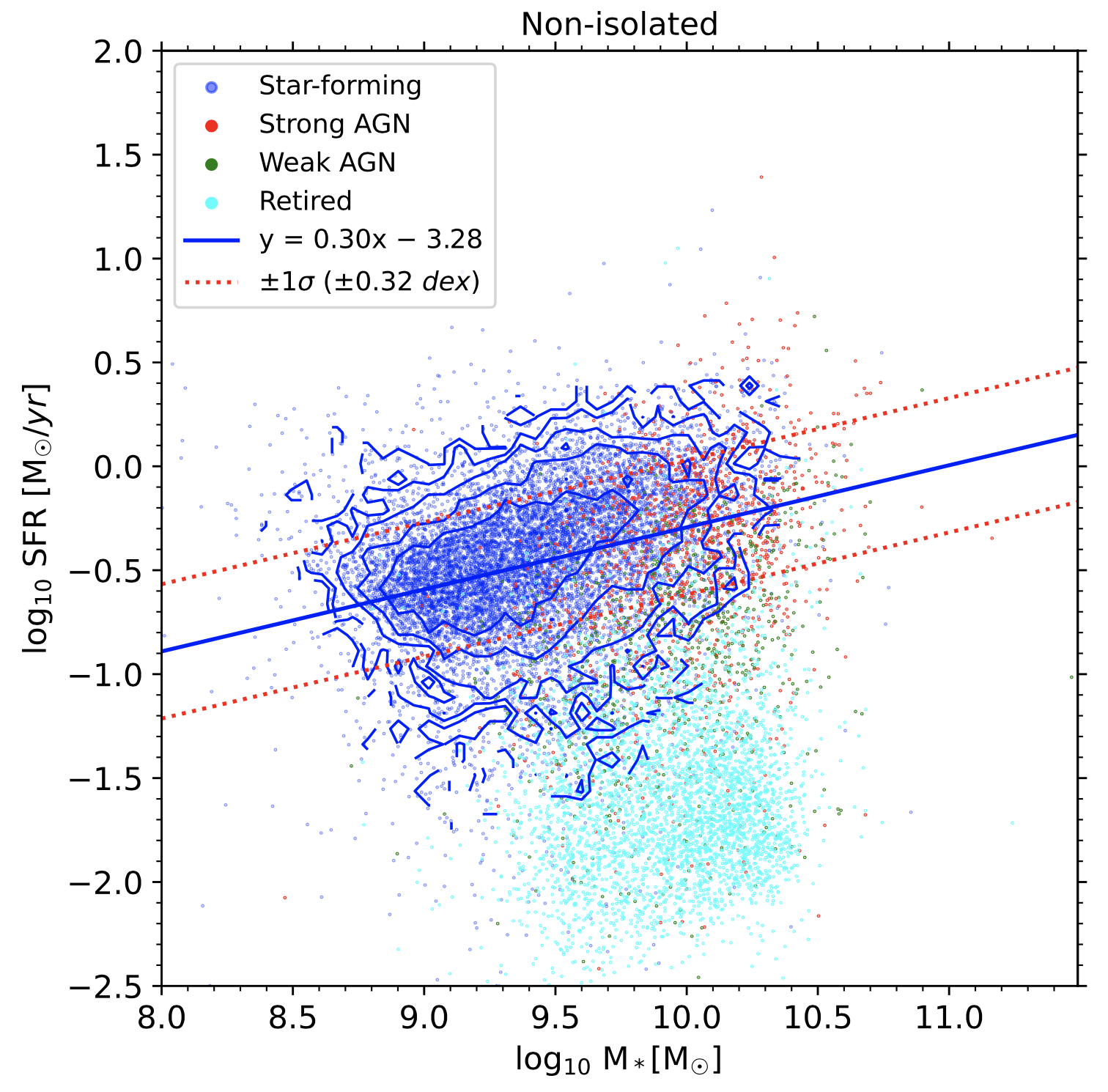}
		}
		\vspace{-0.2cm}
		\caption{Scatter plot showing the SFR as the function of 
			stellar mass  for the luminous volume-limited main galaxy sample 
			(upper panels) and the faint volume-limited main galaxy sample (lower panels)
			with  isolated (left panels) and non-isolated (right panels) SF  
			(blue with contour plot), Strong AGN (red points), weak AGN (green points) 
			and retired (cyan points) galaxies. The width of the MS on each
			diagram correspond to the standard deviation ($\pm 1 \sigma$) representing the scatter around the MS ($ \sim \pm 0.3$ dex).}
		\label{MS}
	\end{figure}

	The equation of the MS for SF galaxies shows a tight relationship 
	between the $\log(\mbox{SFR})$ and $\log(M_\star)$, which has been studied in 
	a number of works \cite{elbaz2007reversal,
		speagle2014highly,leslie2015quenching,daddi2007multiwavelength,yuan2010role,rich2011galaxy,leslie2015quenching,schawinski2007observational}.
	The relationship serves as the tracer on how the stars are formed in relation 
	to the stellar mass within a galaxy. The MS is generally characterized by such 
	a relation as given by
	\begin{equation}
		\log_{10}(\mbox{SFR}) = \beta\, \log_{10}(M_{\star}) + \alpha,
		\label{ms}
	\end{equation}
	where $\beta$ and $\alpha$ are slope and intercept, respectively. Aiming at 
	studying how the environment affects the SFR relative to the $M_{\star}$ 
	for the SF, strong AGN, weak AGN, retired classified sub-samples of galaxies, 
	we generate the equation of the best-fitted line of the SF galaxies (the 
	equation of MS based on Eq.~\eqref{ms}). This best-fitted line is used as a 
	reference to understand the behaviour of SFR relative to $M_{\star}$ for all 
	other sub-samples of galaxies residing in different environments. 
	Fig.~\ref{MS} shows the distributions of SFR with respect to $M_{\star}$ of 
	the SF, strong AGN, weak AGN and retired galaxies for isolated and 
	non-isolated cases along with the corresponding best-fitted MS line of the SF 
	galaxies. The width of the MS of $ \sim \pm\, 0.3$ dex 
	(dashed lines) in the plots of this figure as already stated is selected 
	based on the dispersion of the observed MS. 
	The figures show 
	the positions of galaxies for all sub-samples with respect to the MS and the 
	corresponding numbers of galaxies with their percentages are shown in 
	Table \ref{LMST} for the luminous, and in Table 
	\ref{FMST} for the faint volume-limited sample. 
	\noindent {By performing the regression analysis for the luminous 
		volume limited sample the general equations of the best-fitted line for the 
		isolated and non-isolated galaxies are respectively given by Eqs.~\eqref{SQMI} 
		and~\eqref{SQMNI}. 
		\begin{align}
			\log_{10}(\mbox{SFR}) & = 0.21\pm 0.01\,\log_{10}(M_\star) - 1.87\pm0.10,
			\label{SQMI}\\[5pt]
			\log_{10}(\mbox{SFR}) & = 0.17\pm 0.01\,\log_{10}( M_\star) - 1.48\pm 0.10.
			\label{SQMNI}
		\end{align}
		Similarly, for the faint volume-limited sample the general equations of the 
		the best-fitted line for the isolated and non-isolated galaxies are respectively 
		given by Eqs.~\eqref{SQFI} and \eqref{SQFNI}.
		\begin{align}
			\log_{10}(\mbox{SFR}) & = 0.38\pm0.01\,\log_{10}( M_\star) - 4.02\pm0.10,
			\label{SQFI}\\[5pt]
			\log_{10}(\mbox{SFR}) & = 0.30\pm 0.01\,\log_{10}( M_\star) - 3.28\pm 0.10.
			\label{SQFNI}
		\end{align}
		The associated errors in Eqs.~\eqref{SQMI},~\eqref{SQMNI},~\eqref{SQFI} and~\eqref{SQFNI} are 
		the standard deviations in the slope and intercept. Table \ref{CPMS} indicates the percentage difference
		 between isolated 
		and non-isolated galaxies for the luminous sample above MS, within MS and 
		below MS for SF (column 2), Strong AGN (column 3), Weak AGN (column 4) and 
		retired (column 5) galaxies, while for the faint sample SF (column 6), 
		Strong AGN (column 7), Weak AGN (column 8) and retired (column 9) galaxies, 
		respectively. This table can be used as the tracer on how the positioning of 
		galaxies in the plane of SFR versus $M_\star$ are affected by environment.
		
		\begin{table}[h!]
			\centering
			\caption{Number of galaxies within (MS), above (Above MS), and 
					below (Below MS) the star-forming main sequence for isolated 
					(iso) and non-isolated (nis) environments for the luminous 
					volume-limited sample.}
			\vspace{8pt}
			\setlength{\tabcolsep}{0.34pc}
			\begin{tabular}{ccccccccc}
				\toprule
				\toprule
				\multicolumn{1}{c}{} &
				\multicolumn{2}{c}{{Star-forming (\%)}}&\multicolumn{2}{c}{{Strong AGN (\%)}}&\multicolumn{2}{c}{{Weak AGN(\%)}}&\multicolumn{2}{c}{{Retired (\%)}}
				\\
				\cmidrule(lr){2-3} \cmidrule(lr){4-5}\cmidrule(lr){6-7} \cmidrule(lr){8-9} 
				Position & iso & nis& iso& nis& iso & nis&iso & nis\\
				{(1)} &{(2)}&{(3)}&{(4)}&{(5)}&{(6)}&{(7)}&(8)&(9)\\
				\midrule
				MS & $14480 (79.30)$ & $12883 (75.14)$ & $7682 (58.06)$ & $7819 (54.12)$ & $1362 (26.20)$ & $1522 (22.26)$& $452 (2.12)$ & $645 (1.62)$  \\[1pt]

				Above MS & $1963 (10.75)$ & $2216 (12.93)$ & $1369 (10.35)$ & $ 1918 (13.28)$  & $18 (0.35)$ & $38 (0.56)$ & $10 (0.05)$ & $35(0.09)$\\[1pt]

				Below MS & $1816 (9.95)$ & $2046 (11.93)$ & $4180 (31.59)$ & $4710 (32.60)$ & $3819 (73.46)$ & $5277 (77.18)$& $20887 (97.83)$ & $39127 (98.29) $\\[1pt]
				
				Total	& $18259 (100)$ & $17145 (100)$ & $13231(100)$ & $14447 (100)$  & $5199 (100)$ & $6837 (100)$ & $ 21349 (100)$ & $39807 (100)$\\
				\bottomrule
			\end{tabular}
			\label{LMST}
		\end{table}
		\begin{table}[h!]
			\centering
			\caption{Number of galaxies within (MS), above (Above MS), and 
					below (Below MS) the star-forming main sequence for isolated 
					(iso) and non-isolated (nis) environments for the faint volume 
					limited sample.}
			\vspace{8pt}
			\setlength{\tabcolsep}{0.5pc}
			\begin{tabular}{ccccccccc}
				\toprule
				\toprule
				\multicolumn{1}{c}{} &
				\multicolumn{2}{c}{{Star-forming (\%)}}&\multicolumn{2}{c}{{Strong AGN (\%)}}&\multicolumn{2}{c}{{Weak AGN  (\%)}}&\multicolumn{2}{c}{{Retired (\%)}}
				\\
				\cmidrule(lr){2-3} \cmidrule(lr){4-5}\cmidrule(lr){6-7} \cmidrule(lr){8-9} 
				Position & iso & nis& iso& nis& iso & nis&iso & nis\\
				{(1)} &{(2)}&{(3)}&{(4)}&{(5)}&{(6)}&{(7)}&(8)&(9)\\
				\midrule
				MS & $6416 (76.33)$ & $7018 (69.10) $ & $515 (61.24)$ & $660 (54.77)$ & $ 146 (39.46)$ & $193 (25.70)$& $39 (3.74)$ & $79 (2.11)$  \\[1pt]

				Above MS & $995 (11.84)$ & $1626 (16.01)$ & $70 (8.32)$ & $165 (13.69)$ & $9 (2.43)$ & $8 (1.07)$& $2 (0.19)$ & $8 (0.21)$ \\[1pt]

				Below MS & $995 (11.84)$ & $1513 (14.90)$ & $256 (30.44)$ & $380 (31.54)$ & $215 (58.11)$ & $550 (73.24)$& $1001 (96.07)$ & $3654 (97.68)$  \\[1pt]
				
				Total	& $8406 (100)$ & $10157 (100)$ & $841 (100)$ & $1205 (100)$  & $370 (100)$ & $751 (100)$ & $1042 (100)$ & $3741 (100)$\\
				\bottomrule
			\end{tabular}
			\label{FMST}
		\end{table}
		
		\begin{table}[!h]
			\centering
			\caption{The percentage difference ($\bigtriangleup$ (\%)) of isolated and 
					non-isolated galaxies within MS, above MS and below MS.}
			\vspace{8pt}
			\setlength{\tabcolsep}{0.3pc}
			\scalebox{1}{
				\begin{tabular}{ccccccccccc}
					\toprule
					\toprule
					& \multicolumn{5}{c}{ Luminous $\bigtriangleup$ (\%)} & \multicolumn{5}{c}{Faint $\bigtriangleup$ (\%)} \\
					\cmidrule(lr){2-6} \cmidrule(lr){7-11}  
					Position &  Star-forming & Strong AGN  & Weak AGN & Retired &Average& Star-forming & Strong AGN & Weak AGN  &Retired& Average\\
					{(1)} &{(2)}&{(3)}&{(4)}&{(5)}&{(6)}&{(7)}&(8)&(9)&(10)&(11)\\
					\midrule
					MS &  \num{5.84} & \num{0.88}  &\num{5.55} & \num{17.59}&\num{7.47}& \num{4.48} & \num{12.34} & \num{13.86}& \num{33.90} &\num{16.15}\\[1pt]
					
					Above MS & \num{6.05} & \num{16.70}  & \num{35.71} & \num{55.56}&\num{28.51}& \num{24.07} & \num{40.43}  & \num{5.88}& \num{60.00}&\num{32.60} \\[1pt]		
					
					Below MS & \num{5.96} & \num{5.96}  & \num{16.03} & \num{30.39}&\num{14.59}& \num{20.65}& \num{19.50}&\num{43.79}&\num{56.99}& \num{35.23} \\[1pt]
					\bottomrule
			\end{tabular}}
			\label{CPMS}
		\end{table}
		
		\subsection{ Colour Bimodality}
		Galaxies can be categorized into two groups: those actively forming stars, 
		appearing blue, and those lacking significant star formation, appearing red. 
		The galaxies initially fall into the blue sub-category and then they change 
		gradually to red \cite{gonccalves2012quenching,moustakas2013primus}. It is 
		clear to say that evolution from one category to another must involve 
		processes which quench their rate of forming new stars from the blue cloud 
		passing the intermediate stage (green valley) to the red sequence 
		\cite{faber2007galaxy,hickox2014black}. The factors for this transformation 
		may be due to internal mechanisms like negative feedback from the AGNs and 
		the galaxy environment. To study the effect of the environment on colour, 
		Table \ref{LGVT} shows the number of galaxies with respect to the green 
		valley (GV) for the luminous samples, 
		while Table \ref{FGVT} shows the number of galaxies with respect to the GV 
		for the faint sample, respectively. The position of galaxies on the colour against the stellar mass 
		diagram is shown in Fig~\ref{cm} for the SF, strong AGN, and weak AGN  and 
		retired galaxies respectively. The width of the GV is derived following the 
		Ref.\ \cite{schawinski2014green}, which is obtained from equations, 
		\begin{align}
			u-r & = -\,0.24 + 0.25 \times M_\star,
			\label{gv1}\\[5pt]
			u-r & = -\,0.75 + 0.25 \times M_\star.
			\label{gv2}
		\end{align}
		Here $u$ and $r$ magnitudes were derived from the SDSS database with 
		extinction corrected. Table \ref{CPGV} indicates the percentage 
		change of colour between isolated and non-isolated galaxies for the luminous  
		sample above GV, within GV and below GV for SF (column 2), Strong AGN 
		(column 3), Weak AGN (column 4) and retired (column 5) galaxies, while for the
		faint sample SF (column 6), Strong AGN (column 7), Weak AGN (column 8) and 
		retired (column 9) galaxies, respectively. This table can be used to trace 
		how the positioning of galaxies in the colour against the stellar mass diagram 
		is affected by changing the galaxies' environment.
		
		\begin{figure}[h!]
			\subfigure{
				\includegraphics[width=0.42\linewidth]{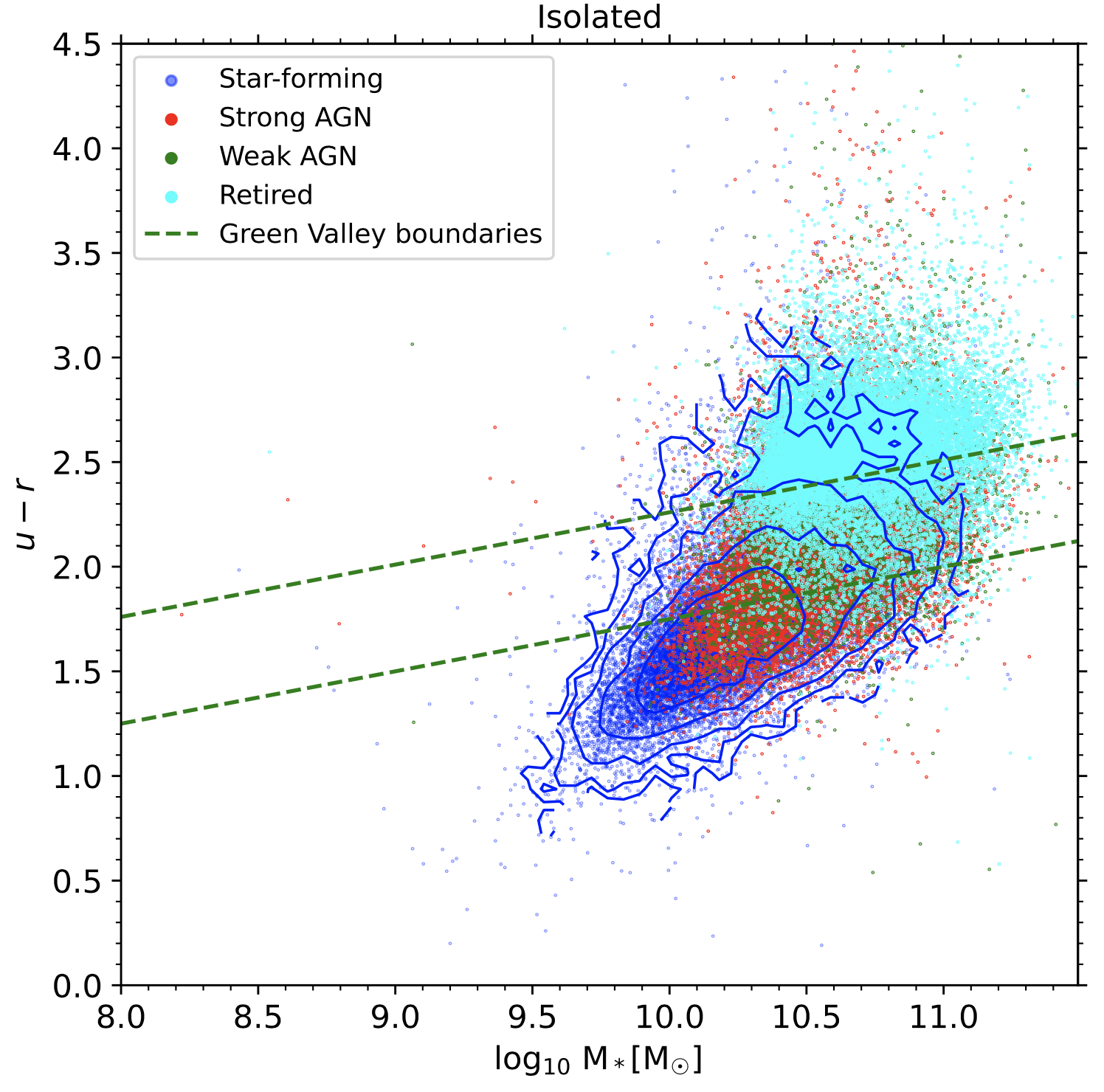}
			}
			\hspace{0.3cm}
			\subfigure{
				\includegraphics[width=0.42\linewidth]{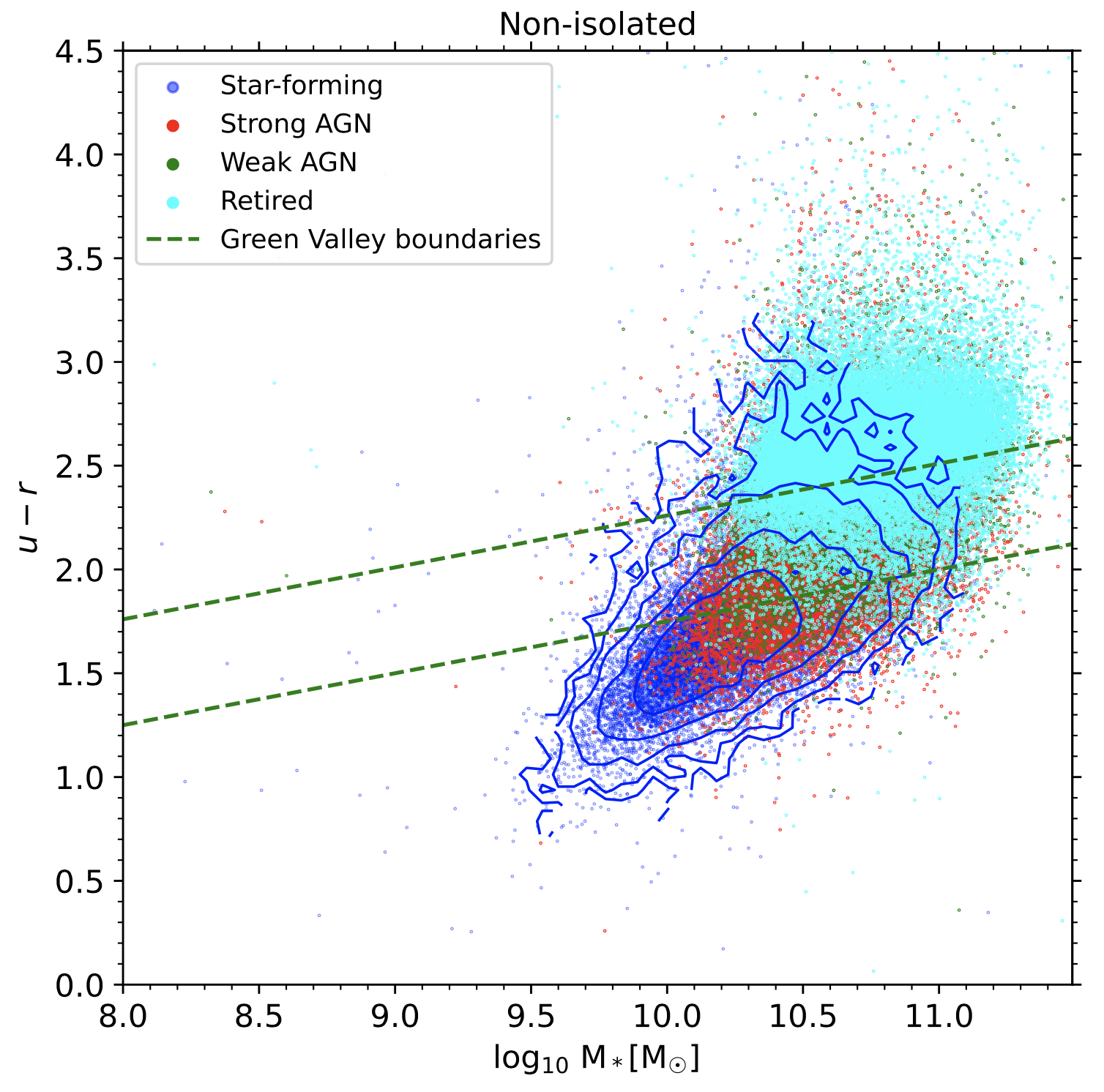}
			}
			\subfigure{
				\includegraphics[width=0.42\linewidth]{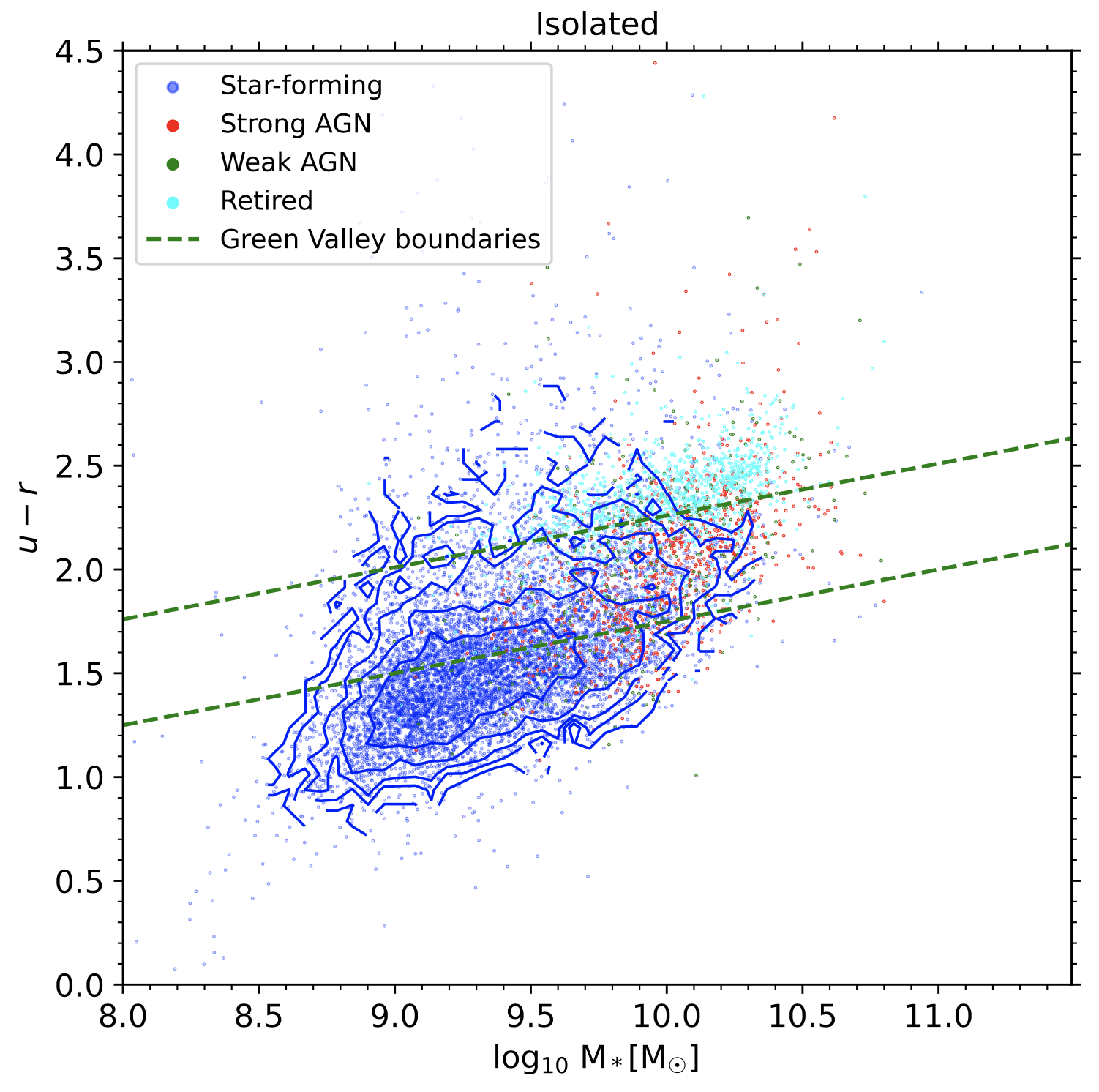}
			}
			\hspace{0.3cm}
			\subfigure{
				\includegraphics[width=0.42\linewidth]{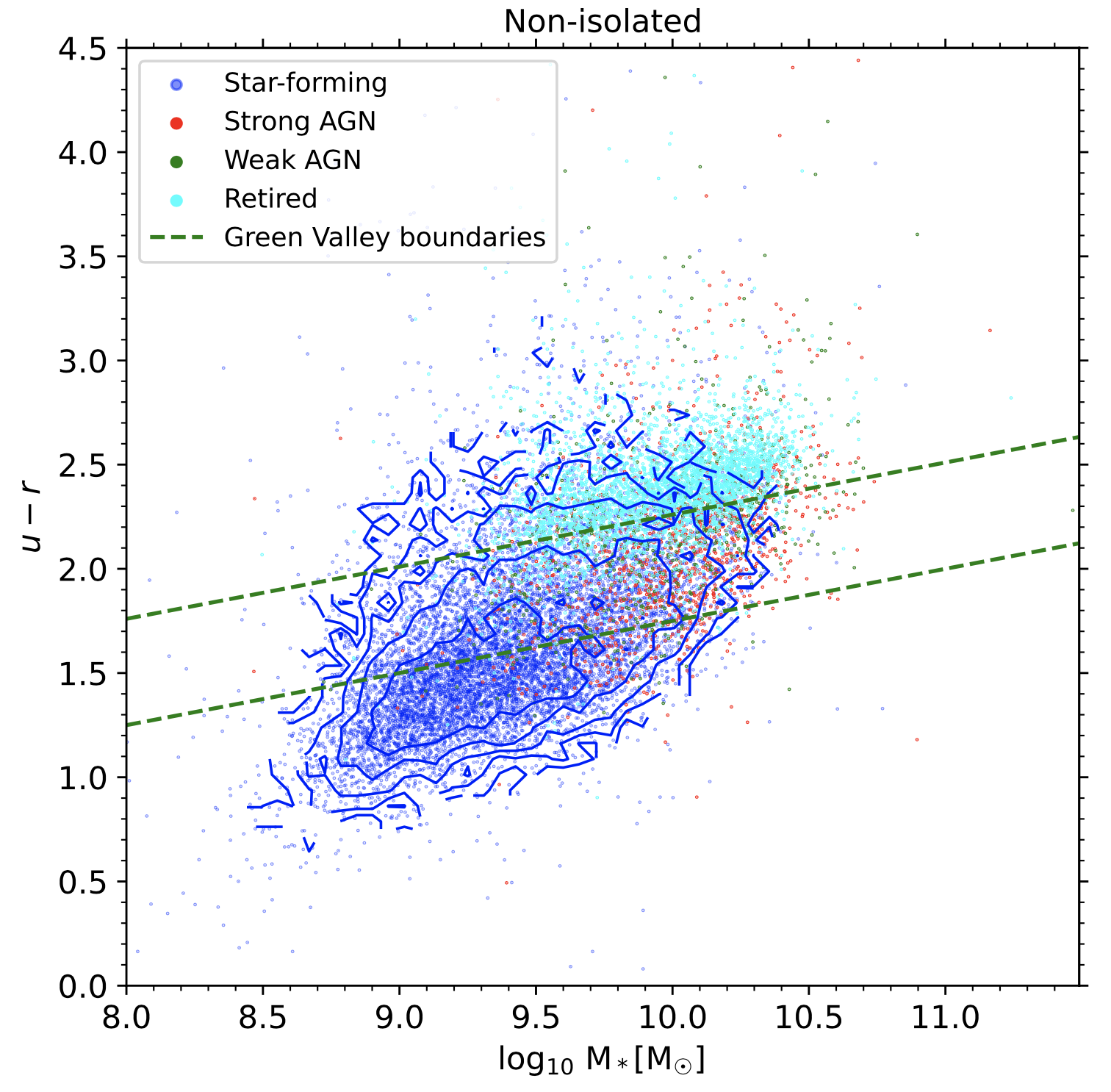}
			}
			\vspace{-0.2cm}
			\caption{Distribution of galaxy's rest-frame colour against 
				stellar mass for the luminous volume-limited main galaxy 
				sample 
				(upper panels) and the faint volume-limited main galaxy 
				sample (lower panels) with isolated (left panels) and 
				non-isolated (right panels) galaxies including SF (blue 
				with contour plots), strong AGN (red points), weak AGN 
				(green points) and retired (cyan points).}
			\label{cm}
		\end{figure}
		
		\begin{table}[ht!]
			\centering
			\caption{Number of galaxies within the green valley (GV), above the
				green valley (Above GV) and below the green valley (Below GV)
				for isolated (iso) and non-isolated (nis) galaxies for the
				luminous volume-limited sample.}
			\vspace{8pt}
			\setlength{\tabcolsep}{0.34pc}
			\begin{tabular}{ccccccccc}
				\toprule
				\toprule
				\multicolumn{1}{c}{} &
				\multicolumn{2}{c}{{Star-forming (\%)}}&\multicolumn{2}{c}{{Strong AGN (\%)}}&\multicolumn{2}{c}{{Weak AGN (\%)}}&\multicolumn{2}{c}{{Retired (\%)}}\\
				\cmidrule(lr){2-3} \cmidrule(lr){4-5}\cmidrule(lr){6-7} \cmidrule(lr){8-9} 
				Position & iso & nis& iso& nis & iso & nis & iso & nis\\
				{(1)} &{(2)}&{(3)}&{(4)}&{(5)}&{(6)}&{(7)}&(8)&(9)\\
				\midrule
				GV & $5023 (27.51)$ & $5360 (31.26)$ & $7302 (55.19)$ & $8015 (55.48)$ & $2854 (54.90)$ &  $3560 (52.07)$ & $5606 (26.26)$ & $8819 (22.15)$ \\[1pt]
				Above GV & $881 (4.83)$ & $1085 (6.33)$ & $1872 (14.15)$ & $2443 (16.91)$ & $1494 (28.74)$  & $2512 (36.74)$ & $15123 (70.84)$&$30303 (76.12)$\\[1pt]
				Below GV & $12355 (67.67)$ & $10700 (62.41)$ & $4057 (30.66)$ & $3989 (27.61)$ & $851 (16.37)$ & $765 (11.19)$ & $620 (2.90)$ & $685 (1.72)$ \\[1pt]
				Total	& $18259 (100)$ & $17145 (100)$ & $13231 (100)$ & $14447 (100)$ & $5199 (100)$ & $6837 (100)$ & $21349 (100)$ & $39807 (100)$ \\
				\bottomrule
			\end{tabular}
			\label{LGVT}
		\end{table}
		
		\begin{table}[ht!]
			\centering
			\caption{Number of galaxies within the green valley (GV), above the
				green valley (Above GV) and below the green valley (Below GV)
				for isolated (iso) and non-isolated (nis) galaxies for the 
				faint volume-limited sample.}
			\vspace{8pt}
			\setlength{\tabcolsep}{0.5pc}
			\begin{tabular}{ccccccccc}
				\toprule
				\toprule
				\multicolumn{1}{c}{} &
				\multicolumn{2}{c}{{Star-forming (\%)}}&\multicolumn{2}{c}{{Strong AGN (\%)}}&\multicolumn{2}{c}{{Weak AGN (\%)}}&\multicolumn{2}{c}{{Retired (\%)}}\\
				\cmidrule(lr){2-3} \cmidrule(lr){4-5}\cmidrule(lr){6-7} \cmidrule(lr){8-9} 
				Position & iso & nis& iso& nis & iso & nis & iso & nis\\
				{(1)} &{(2)}&{(3)}&{(4)}&{(5)}&{(6)}&{(7)}&(8)&(9)\\
				\midrule
				
				GV & $2863 (34.06)$ & $3874 (38.14)$ & $513 (61.00)$ & $710 (58.92)$ & $202 (54.59)$ &  $374 (49.80)$ & $245 (23.51)$ & $794 (21.22)$ \\[1pt]
				Above GV & $615 (7.32)$ & $1026 (10.10)$ & $171 (20.33)$ & $334 (27.72)$ & $109 (29.46)$ & $328 (43.68)$ & $774 (74.28)$ & $2911 (77.82)$\\[1pt]
				Below GV & $4928 (58.62)$ & $5257 (51.76)$ & $157 (18.67)$ & $161 (13.36)$ & $59 (15.95)$ & $49 (6.52)$ & $23 (2.21)$ & $36 (0.96)$ \\[1pt]
				Total	& $8406 (100)$ & $10157 (100)$ & $841 (100)$ & $1205 (100)$ & $370 (100)$ & $751 (100)$ & $1042 (100)$ & $3741 (100)$ \\
				\bottomrule
			\end{tabular}
			\label{FGVT}
		\end{table}
		\begin{table}[!h]
			\centering
			\caption{Percentage difference ($\bigtriangleup$ (\%)) of isolated and 
					non-isolated galaxies within the green valley (GV), above the 
					green valley (Above GV) and below the green valley (below GV) 
					for luminous and faint samples. }
			\vspace{8pt}
			\setlength{\tabcolsep}{0.3pc}
			\scalebox{1}{
				\begin{tabular}{ccccccccccc}
					\toprule
					\toprule
					& \multicolumn{5}{c}{ Luminous $\bigtriangleup$ (\%)} & \multicolumn{5}{c}{Faint $\bigtriangleup$ (\%)} \\
					\cmidrule(lr){2-6} \cmidrule(lr){7-11}  
					Position &  Star-forming & Strong AGN & weak AGN & Retired &Average& Star-forming & Strong AGN & weak AGN& Retired&Average \\
					{(1)} &{(2)}&{(3)}&{(4)}&{(5)}&{(6)}&{(7)}&(8)&(9)&(10)&(11)\\
					\midrule
					GV &  \num{3.25} & \num{4.65} & \num{11.01}&\num{22.27}&\num{10.30}& \num{15.01} & \num{16.11} & \num{29.86}& \num{52.84}&\num{28.46}\\[1pt]
					Above GV & \num{10.38} & \num{13.23} & \num{25.41}& \num{33.42} &\num{20.61}& \num{25.05} & \num{32.28} & \num{50.11}& \num{57.99} &\num{41.36}\\[1pt]		
					Below GV & \num{7.18} & \num{0.85}& \num{4.98}& \num{9.96}&\num{5.74}& \num{3.23} & \num{1.26}&\num{9.26}&\num{22.03} & \num{8.95}\\[1pt]
					\bottomrule
			\end{tabular}}
			\label{CPGV}
		\end{table}
		
		\section{Discussion}
		\label{secV}
		From Fig~\ref{WHAN} it is observed that the fraction of SF galaxies 
		for the faint volume-limited sample is always larger than the luminous 
		volume-limited sample for both isolated and non-isolated galaxies. For luminous 
		galaxies, the fraction of RGs is larger than SF, strong AGN, and weak AGN. 
		while for the faint volume-limited sample the fraction of star-forming 
		galaxies are larger than strong AGN, weak AGN, and retired galaxies. 
		In the luminous volume-limited sample, the difference in the fraction of galaxies 
		between isolated and non-isolated galaxies are $3.15\%$, $4.39\%$, $13.61$, 
		$30.18\%$ for SF, strong AGN, weak AGN, and retired galaxies, respectively. In 
		the faint sample, the difference in fractions are $9.43\%$, $17.79\%$, $33.99\%$, 
		$56.43\%$ for SF, strong AGN, weak AGN, and retired galaxies, respectively.  
		Again from Fig~\ref{WHAN}, we observe that the fraction of isolated strong AGN 
		for both luminous and faint volume-limited samples is always larger 
		than the non-isolated sample. Since the non-isolated environment is observed to 
		be an area of high density our study agrees with 
		Ref.\ \cite{deng2012environmental}, which studied the AGNs and observed that 
		the fraction of low-density AGNs is larger than the sample of higher-density 
		AGNs.
		
		From Fig~\ref{MS}, the difference of $0.04$ dex in slope, and $0.39$ 
		dex in intercept between the isolated and non-isolated luminous volume-limited 
		samples are observed, while for the faint volume-limited sample a difference of 
		$0.08$ dex in slope and $0.74$ dex in intercept is observed. Using two samples 
		t-statistics test P-values of $\num{7.19e-04}$, $\num{1.16e-03}$ for the luminous 
		sample and $\num{8.78e-10}$, $\num{3.97e-09}$ for the faint sample in slope, 
		intercept respectively, were obtained indicating that the observed differences
		in slope and intercept for both luminous and faint sample are statistically 
		significant. From the equations of star formation MS, it is observed that the 
		slope of the isolated galaxies is greater than non-isolated galaxies. This 
		originated from the fact that the SFR decreases for non-isolated galaxies 
		while stellar mass increases when the galaxies are with companions. 
		These results support the findings of Refs.\ \cite{lewis20022df, 
			gomez2003galaxy, tanaka2004environmental, elbaz2007reversal, cooper2008deep2, 
			patel2009dependence,mountrichas2023relation,mountrichas2024probing}, regarding 
		fluctuations in  SFR with environment. On the other hand, the luminous 
		volume-limited samples possess lower slopes than the faint volume-limited samples for 
		both isolated and non-isolated galaxies this indicates that the faint galaxies 
		possess higher SFR and lower masses than the luminous volume-limited sample. Again 
		from Fig~\ref{MS} the stellar mass is observed to increase following the 
		sequence SF, strong AGN, weak AGN and retired galaxies. The SFR decreases 
		in the same sequence, agreeing with the findings by Ref.\ \cite{deng2012some} 
		which observed that galaxies hosting AGNs are preferentially more massive and 
		have low SFR. The study is consistent with the results from 
		Ref.~\cite{sanchez2018sdss} which presented a characterization of the primary 
		characteristics of a sample of 98 AGN host galaxies, both type-II and 
		type-I. They found that, on average, AGN hosts are more massive and more 
		compact than SF galaxies, compared with those of $\simeq 2700$ non-active 
		galaxies observed by the MaNGA survey. 
		
		The increase of $0.17$ dex and $2.15$ dex in slope and intercept 
		respectively is observed between the isolated luminous and the isolated faint
		volume-limited samples, while the increase of $0.13$ dex in slope and $1.8$ dex 
		in intercept is observed between non-isolated luminous 
		and faint volume-limited samples indicating that luminosity is another very 
		an important factor shaping the MS, where the effect is significant for the 
		isolated than the non-isolated galaxies. From Table \ref{LMST} and \ref{FMST}, 
		it is observed that there is a difference in fractions between isolated and 
		non-isolated galaxies for SF, strong AGN, weak AGN and retired galaxies 
		as shown by Table \ref{CPMS}, where for the luminous sample the differences of the
		number of galaxies within, above and below the main sequence are respectively 
		7.47\%, 28.51\%, 14.59\%, while for the faint sample the differences are 
		16.15\%, 32.60\%, 35.23\% on average, respectively. This differences 
		produce an average P-values of $\num{8.37e-10}$, $\num{2.48e-03}$, 
		$\num{4.13e-03}$, $\num{3.44e-03}$ for the luminous sample and 
		$\num{4.89e-10}$, $\num{2.12e-03}$, $\num{4.44e-03}$, $\num{3.37e-03}$ 
		for the faint sample under a two-sample chi-square statistical test ($\chi^2$) 
		for SF, strong AGN, weak AGN, and retired galaxies, respectively. These P 
		values are much less than the standard value in statistics ($0.05$), 
		indicating that the differences are statistically significant supporting the 
		evidence from  Refs.\ \cite{cristello2024investigating, mountrichas2022comparisona,mountrichas2022comparisonb,riffel2023mapping}, 
		regarding the positioning of galaxies with respect to the main sequence.
		
		From Table \ref{LGVT} and \ref{FGVT}, it is observed that there is a 
		difference in fraction between isolated and non-isolated galaxies for SF, 
		strong AGN, weak AGN and retired galaxies as shown by Table \ref{CPGV}, 
		indicating that position of galaxies with respect to the green valley is 
		affected by the environment for both luminous and faint volume galaxy samples.  
		The difference of the number of galaxies within, above and below the 
		green valley by 10.30\%, 20.61\%, 5.74\% for the luminous sample while for 
		the faint sample by  28.46\%, 41.36\%, 8.95\% on average, respectively is 
		observed. These differences produce an average P-value of 
		$\num{2.59e-10}$, $\num{2.12e-03}$, $\num{7.40e-04}$, $\num{2.12e-22}$ for the 
		luminous sample and $\num{3.13e-09}$, $\num{1.24e-03}$, $\num{4.93e-04}$, 
		$\num{4.79e-04}$ for the faint sample under a two-sample chi-square statistical 
		test ($\chi^2$) for SF, strong AGN, weak AGN, and retired galaxies 
		respectively. These P-values indicate that the differences are statistically 
		significant which means there is a significant difference in the number of 
		galaxies in blue cloud, green valley and the red sequence between isolated and 
		non-isolated environments. Hence the galaxy environment influences colour 
		bimodality. The u-r colour against $M_\star$ diagrams in Fig.~\ref{cm}, 
		reinforces the relation with respect to the MS in Eq~\eqref{ms}, 
		the strong AGN galaxies are more redder and more massive than the SF, the weak AGN 
		galaxies are more massive and redder than strong AGN, and retired have the 
		highest stellar mass and most redder. These observations are consistent with 
		the results of Ref.\ \cite{leslie2015quenching}, which found that the SF was 
		blue in colour and had lower masses. The larger fraction of galaxies within 
		and above the GV in Tables \ref{FGVT} and \ref{LGVT} 
		show that the non-isolated environment accelerates the rate of change in 
		SFR in comparison to the isolated environment. 
		Furthermore, the lower fraction of the number of galaxies below the green valley 
		for non-isolated galaxies when compared to isolated galaxies indicates the 
		decrease of SFR for non-isolated galaxies. From Fig.~\ref{cm}, it is observed 
		that the SF galaxies form a blue cloud since most of the galaxies are below 
		the green valley as revealed by Table \ref{LGVT} and \ref{FGVT}, where 
		$\sim\!60\%$ of SF are below the main sequence. Strong AGN are in a green 
		valley as most of the galaxies lie within the demarcation lines 
		($\sim\! 58\%$) on average. Furthermore, it indicates that most of weak AGN 
		galaxies galaxies are within the green valley ($\sim\!53\%$) on average and 
		retired galaxies are above the green valley ($\sim 75\%$) on average.  These
		results are consistent with Ref.\ \cite{lacerda2020galaxies}, wherein it
		is observed that the AGN hosts are found in transitory parts (i.e., 
		green-valley) in almost all analysed properties which present bimodal 
		distributions, using 867 galaxies extracted from the extended Calar-Alto 
		Legacy Integral Field spectroscopy Area (eCALIFA) \cite{sanchez2016califa}. Similar to this, 
		the study by Ref.\ \cite{sanchez2018sdss} found that AGNs are situated in the 
		`green valley', the transition area between SF and 
		non-SF galaxies, and they hypothesized that this star formation is in the 
		process of halting or quenching. 
		
		\section{Summary and Conclusion}
		\label{secVI}
		For the first time we use the friend-of-friend method as 
		detailed in Ref.\ \cite{tempel2017merging} to study the dependence of star 
		formation MS and colour bimodality on the environment. Using the flux 
		limited sample from Sloan Digital Sky Survey as detailed in 
		Refs.\ \cite{eisenstein2011sdss,alam2015eleventh}, we constructed the 
		volume-limited samples of luminous and faint galaxies. The galaxies were 
		classified into two systems within a specific radius. The galaxies with no 
		companion are termed isolated while the one with more than one neighbour 
		in a system are termed non-isolated. Using the WHAN diagram we classified the 
		galaxies into SF, strong AGN, weak AGN, and retired galaxies. We used the 
		stellar mass and SFR retrieved from MPA-JHU obtained using 
		the methods outlined in Ref.\ \cite{brinchmann2004physical,tremonti2004origin},
		to include the strong AGN, weak AGN and retired galaxies on the local star 
		formation MS and colour, stellar mass diagram, then we study the influence of 
		isolated and non-isolated environments on the star formation MS, and colour 
		bimodality. The friends-of-friends method produced a consistent 
		results with the previous studies obtained by different methods \cite{deng2012environmental,speagle2014highly,leslie2015quenching,sanchez2018sdss,lacerda2020galaxies,mountrichas2022comparisona,mountrichas2022comparisonb,sanchez2018sdss,riffel2023mapping,cristello2024investigating,mountrichas2023relation,mountrichas2024probing}. 
		Apart from that our study revealed the following:
		
		\begin{itemize}
			\item The decrease of the slope for the MS of SF galaxies by $0.04$
			dex and intercept by $0.39$ dex for the luminous volume limited 
			sample, while for the faint volume-limited sample a decrease of $0.08$ dex in 
			slope, and $0.74$ dex in intercept are observed between isolated and 
			non-isolated galaxies. The differences produced a  P-value of  $\num{7.19e-04}$, $\num{1.16e-03}$ for 
			the luminous sample and $\num{8.78e-10}$, $\num{3.97e-09}$ for the faint sample in 
			slope and intercept, respectively proving that the difference is 
			statistically significant implying that the environment is among the factors which contribute to the
			shaping of the star formation main sequence.
			\item A significant difference in the number of galaxies between isolated and non-isolated environment within, above and  below the main sequence by 7.47\%, 28.51\%, 14.59\% for the luminous and  by 16.15\%, 
32.60\%, 35.23\% for faint samples on average, respectively are observed 
indicating that the environment influence the positioning of galaxies along 
the star formation MS.
			\item For colour bimodality  a significant difference 
in the number of galaxies within, above and below the green valley by 
10.30\%, 20.61\%, 5.74\% for the luminous sample and by 28.46\%, 41.36\%, 
8.95\% for the faint sample on average, respectively between isolated and 
non-isolated galaxies are observed which implies that colour bimodality is 
influenced by the galaxy's environment.
		\end{itemize}
		In the next study (in preparation), as the follow-up of this work, we aim to investigate if ageing and 
		quenching processes are influenced by the environment using the same method.
		
		\section*{Acknowledgements} PP acknowledges support from The Government of 
		Tanzania through the India Embassy, Mbeya University of Science and Technology 
		(MUST) for Funding. UDG is thankful to the Inter-University Centre for 
		Astronomy and Astrophysics (IUCAA), Pune, India for the Visiting 
		Associateship of the institute. Funding for SDSS- III has been provided by 
		the Alfred P. Sloan Foundation, the Participating Institutions, the National Science Foundation, 
		and the U.S. Department of Energy Office of Science. 
		The SDSS-III website is http://www.sdss3.org/. SDSS-III is managed by the 
		Astrophysical Research Consortium for the Participating Institutions of the 
		SDSS-III Collaboration including the University of Arizona, the Brazilian Participation Group, 
		Brookhaven National Laboratory, Carnegie Mellon University, the University of Florida, 
		the French Participation Group, the German Participation Group, Harvard University, 
		the Instituto de Astrofisica de Canarias, the Michigan State/Notre Dame/JINA Participation Group, 
		Johns Hop- kins University, Lawrence Berkeley National Laboratory, Max Planck Institute for Astrophysics, 
		Max Planck Institute for Extraterrestrial Physics, New Mexico State University, 
		New York University, Ohio State University, Pennsylvania State University, 
		University of Portsmouth, Princeton University, the Spanish Participation Group, University of Tokyo, 
		the University of Utah, Vanderbilt University, the University of Virginia, the University of Washington, and Yale University.
		
	\end{document}